\documentclass[doublespacing]{elsart}

\usepackage{graphicx}  
\usepackage{amssymb}   
\usepackage{bm}        

\newcommand{\const}{\mathrm{const}}
\newcommand{\emit}{\varepsilon}

\begin{document}

\begin{frontmatter}


 
\title{ Space-charge transport limits of ion beams
in periodic quadrupole focusing channels }


\author{Steven M. Lund}
\address{Lawrence Livermore National Laboratory, Livermore, CA 94550, USA}
\ead{SMLund@llnl.gov}
\author{Sugreev R. Chawla}
\address{Lawrence Berkeley National Laboratory, Berkeley, CA 94720, USA}

\date{February 23, 2006}

\begin{abstract}

It has been empirically observed in both experiments and
particle-in-cell simulations that space-charge-dominated beams
suffer strong growth in statistical phase-space area (degraded 
quality) and particle losses in
alternating gradient quadrupole transport
channels when the undepressed phase advance $\sigma_0$ increases
beyond about $85^\circ$ per lattice period.  Although this criterion
has been used extensively in practical designs of strong focusing intense 
beam transport lattices, the origin of the limit has not been understood.   
We propose a mechanism for the transport limit
resulting from classes of halo particle resonances near the core of
the beam that allow near-edge particles to rapidly increase
in oscillation amplitude when
the space-charge intensity and the flutter of the matched beam envelope
are both sufficiently large.  When coupled with a diffuse beam edge and/or 
perturbations internal to the beam core that can drive particles outside 
the edge, this mechanism can result in large and rapid halo-driven increases
in the statistical phase-space area of the beam,
lost particles, and degraded transport.  A core-particle model 
is applied to parametrically analyze this process.  Extensive self-consistent
particle in cell simulations are employed to better quantify space-charge 
limit and verify core-particle model predictions.

\end{abstract}
 
\begin{keyword}
intense beam \sep space charge \sep emittance growth \sep simulation 
\PACS 29.27.Bd \sep 41.75.-i \sep 52.59.Sa \sep 52.27.Jt 
\end{keyword}


\end{frontmatter}

\section{Introduction }

The maximum transportable current density of an ion beam with high
space-charge intensity propagating in a periodic focusing lattice is
a problem of practical 
importance\cite{Tiefenback-1985,Tiefenback-1986}. Accelerator
applications such as Heavy Ion Fusion (HIF), High Energy Density
Physics (HEDP), and transmutation of nuclear waste demand a large flux of
particles on target.  A limit to the maximum 
current density can result from a variety
of factors: instability of low-order moments of the beam describing
the centroid and envelope, instability of higher order collective
modes internal to the beam, growth in statistical phase-space area 
(rms emittance growth), excessive halo generation, and
species contamination associated with issues such as the electron
cloud problem. Simulations were first used to analyze the maximum 
current density transportable in quadrupole 
channels\cite{Haber-misc,Struckmeier-1984} and provided 
guidance beyond initial heuristic estimates\cite{Maschke-1972}. 
Experiments later obtained results consistent with 
simulations\cite{Tiefenback-1985,Tiefenback-1986}.   

The present work describes a promising new approach toward predicting 
the maximum transportable current density in a periodic quadrupole lattice 
due to intrinsic space-charge limits\cite{Lund-2005}.   Previous 
studies to predict space-charge related transport limits in the absence of
focusing errors and species contamination have not proved fully 
successful beyond a moment level description of low-order beam 
instabilities.  Although moment-based centroid and envelope descriptions
reliably predict regions of parametric instability where machines
cannot operate\cite{Reiser-1994,Lund-2004}, such models are 
overly optimistic when compared to simulations and experiments which 
observe degraded transport due to emittance growth and particle losses 
where the moment models predict 
stability\cite{Tiefenback-1985,Tiefenback-1986,Haber-misc,Struckmeier-1984}.
On the other hand, higher-order collective mode theories 
based on the equilibrium KV distribution\cite{Hofmann-1983} predict 
broad parametric regions of instability where stability is observed in 
simulations with more realistic 
distributions\cite{Haber-misc,Struckmeier-1984} and in 
experiment\cite{Tiefenback-1985,Tiefenback-1986}.  The 
space-charge limit model proposed is 
based on particles oscillating outside, but near the beam edge 
exchanging energy with the oscillating space-charge field of a 
envelope matched beam core leading to 
increased particle oscillation amplitude, emittance blow up, 
and particle losses.  This model can be applied to a wide range of 
matched core distributions and does not require 
an equilibrium core -- which circumvents the practical problem of 
no smooth core equilibrium distribution being known.   
The increased understanding the origin of the observed limits 
obtained promises more reliable 
design of optimal intense beam transport channels. 

We denote the phase advance of particles oscillating in a periodic
focusing lattice in the presence and absence of beam space-charge by
$\sigma$ and $\sigma_0$ (both measured in degrees per lattice
period)\cite{Lund-2004,Reiser-1994}.  The undepressed 
phase-advance $\sigma_0$ provides a measure of the 
strength of the linear applied focusing forces of the
lattice that is relatively insensitive to the details of the
lattice. $\sigma_0$ is generally made as 
large as beam stability will allow --
because stronger focusing results in smaller beam cross-sectional
area leading to smaller, more economical accelerator structures.
$\sigma$ can be unambiguously defined by an rms
equivalent, matched KV equilibrium beam\cite{Lund-2004,Reiser-1994} 
where all particles internal to the beam have the same 
phase advance. The ratio $\sigma/\sigma_0$ is a normalized 
measure of relative space-charge strength with $\sigma/\sigma_0 \rightarrow 1$
corresponding to a warm beam with zero space-charge forces and
$\sigma/\sigma_0 \rightarrow 0$ corresponding to a cold beam with 
maximum space-charge forces.   The maximum possible 
current density for a specified beam line-charge density will occur 
when $\sigma/\sigma_0$ is as small as possible.

\begin{figure}[htb]
\centering
\includegraphics*[width=80mm]{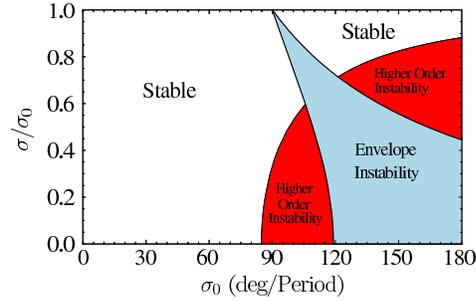}
\caption{
(color online) Beam stability regions in a FODO quadrupole lattice.  
}
\label{Fig:transport-params}
\end{figure}

Neglecting image charge effects, single particle and beam centroid
oscillations are stable if 
$\sigma_0 < 180^\circ$\cite{Reiser-1994}. The
parameter space $\sigma_0 \in (0,180^\circ )$ and
$\sigma/\sigma_0 \in (0,1)$ can be regarded as potential machine 
operating points.  Envelope models predict well understood 
bands of strong parametric instability when 
$\sigma_0 > 90^\circ$ and $\sigma/\sigma_0 < 1$\cite{Lund-2004}.
The parameter region excluded by envelope instabilities 
for FODO quadrupole transport is indicated
(in blue) on Fig.~\ref{Fig:transport-params}.

Considerations beyond centroid and envelope instabilities
exclude further regions of $\sigma_0$--$\sigma$ parameter space.
Transportable current limits based on preservation of beam
statistical emittance and suppression of particle losses for a
matched beam propagating 
in a periodic FODO lattice of 84 electric quadrupoles 
were experimentally studied by Tiefenback at 
LBNL\cite{Tiefenback-1985,Tiefenback-1986}. 
It was found empirically that transport was stable 
(i.e., statistical emittance growth and particle losses below 
measurement thresholds) when 
%
\begin{equation}
\sigma_0^2 - \sigma^2 < \frac{1}{2}( 120^\circ )^2 .
\label{Eq:Tiefenback-limit}
\end{equation}
The additional parameter region this criterion excludes for machine 
operation (partially overlapping the envelope band) 
is indicated (in red) on Fig.~\ref{Fig:transport-params}.  For 
high space-charge intensity
with $\sigma/\sigma_0 < 0.5$, this limit is
more important than the envelope instability band because it is
encountered first when approaching from low $\sigma_0$.  The stability
bound~(\ref{Eq:Tiefenback-limit}) has been applied
by simply requiring that
$\sigma_0 < 120^\circ/\sqrt{2} \simeq 85^\circ$.  
It is observed that
transport becomes more sensitive to errors near the boundary of stability.
%

%
\section{Particle-in-Cell simulations}

Self-consistent electrostatic 
particle-in-cell (PIC) simulations have been carried
out for a variety of initial beam distributions launched in a FODO
quadrupole transport channel with 50\% quadrupole occupancy ($\eta = 1/2$)
and linear, piecewise-constant quadrupole forces.    
The transverse slice module of the WARP code\cite{Grote-1998} is 
employed to advance an initial transverse distribution with zero axial 
velocity spread.  Applied focusing forces are adjusted
for specified $\sigma_0$.  Currents are adjusted
for specified $\sigma/\sigma_0$ using fixed rms emittances 
($\emit_x = \emit_y = 50$ mm-mrad).  Numerical 
parameters are set for high resolution 
($\geq 100$ radial zones across the beam core 
on a square mesh and $\geq 100$ residence 
corrected axial steps per lattice period) and good 
statistics ($\geq 400$ particles per cell).   A cylindrical beam pipe is 
large enough to suppress particle losses and image charge effects. 
Simulation results are in qualitative agreement with
Eq.~(\ref{Eq:Tiefenback-limit}) for a wide variety of initial distribution 
functions.  Initial distributions employed are 
rms matched transforms of continuous focusing 
equilibrium Waterbag and thermal distributions\cite{Reiser-1994}, 
KV, or semi-Gaussian distributions.     
This contrasts earlier work where Waterbag and Gaussian loads did not 
include space-charge screening effects and were far from 
initial force-balance\cite{Struckmeier-1984}.  Only the initial KV 
load employed is a true equilibrium of the periodic focusing channel.  
No exact, smooth equilibrium distributions are presently known for 
periodic focusing channels.   

\begin{figure}[htb]
\centering
\includegraphics*[width=80mm]{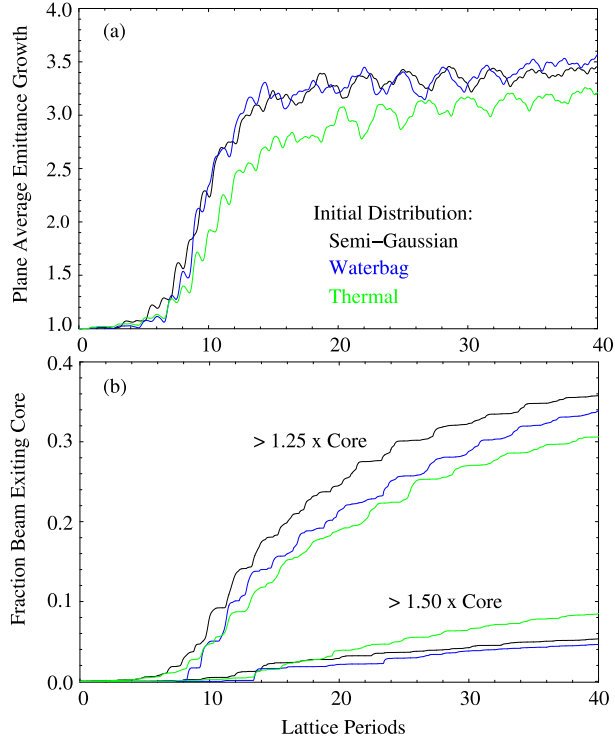}
\caption{
(color online) PIC simulations of (a) the plane-averaged emittance growth
for different initial distributions in a FODO quadrupole channel, and 
(b) the fraction of the beam distribution evolving outside the core.  
($\sigma_0 = 100^\circ$, $\eta = 0.5$, $L_p = 0.5$ m,
$\sigma/\sigma_0 = 0.2$, and $\emit = 50$ mm-mrad).
}
\label{Fig:sim-emit-evolve}
\end{figure}

Parameters to the right of the
stability bound [Eq.~(\ref{Eq:Tiefenback-limit})] and to the left 
of the envelope instability band lead to
statistical (rms) emittance growth and particle losses.  $x$- and $y$-plane 
average emittance [$(\emit_x + \emit_y)/2$]
growth can be rapid and large as illustrated in
Fig.~\ref{Fig:sim-emit-evolve}(a) for a focusing channel with 
$\sigma_0 = 100^\circ$ and three initial distributions: semi-Gaussian, 
waterbag ``equilibrium'', and thermal ``equilibrium.''
Much of this emittance growth can be traced to particles that evolve 
significantly outside the beam core as evident from 
Fig.~\ref{Fig:sim-emit-evolve}(b) which shows the fraction of beam 
particles which evolve (at one or more points) more than 1.25 and 1.5 
times the statistical beam edge radius (i.e., 
$\sqrt{x^2/r_x^2 + y^2/r_y^2} > 1.25$, $1.5$ with 
$r_x = 2\langle x^2 \rangle^{1/2}$ and 
$r_x = 2\langle x^2 \rangle^{1/2}$ calculated from the evolving 
distribution).  The similarity of the results 
for the three very different non-equilibrium
distributions shows that processes degrading the beam are
relatively insensitive to the form of the initial distribution in deeply 
unstable parameter regions.  Laboratory beams are born off a
source (injector) and subsequently manipulated to match into a transport
channel and are unlikely to be any equilibrium form.  
More detailed analysis of the simulation results
show that initial beam distortions leading to the statistical emittance
growth are primarily near the edge of the beam and subsequently act
to strongly perturb the core.   Core perturbations are observed in
both the local density and temperature profiles.  These perturbations
typically lack elliptical symmetry and rapidly oscillate into the core
with excursions larger near the beam edge.  Beam envelope matches are not 
significantly degraded in the initial stages of instability.  

\begin{figure}[htb]
\centering
\includegraphics*[width=80mm]{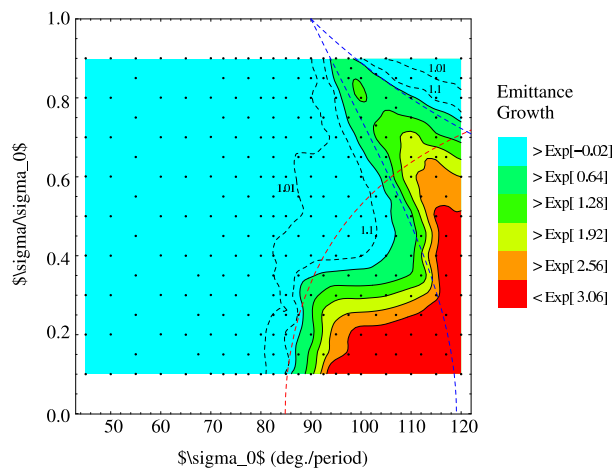}
\caption{
(color online) Contours of emittance growth for an initial semi-Gaussian 
distribution in a FODO quadrupole channel.  
($\eta = 0.5$, $L_p = 0.5$ m, and $\emit = 50$ mm-mrad).
}
\label{Fig:sim-emit-growth}
\end{figure}

A large number of PIC simulations were carried out to better quantify 
parametric regions of instability.  Plane averaged emittance growth contours 
in $\sigma_0$ and $\sigma/\sigma_0$ are shown in 
Fig.~\ref{Fig:sim-emit-growth} for an initial semi-Gaussian distribution.  
Irregular grid simulation points are indicated with dots.  All simulations 
are advanced for six undepressed betatron periods, which is sufficient  
for saturation in strongly unstable regimes.  Near the stability boundary, 
emittance growth slows and growth factors increase with longer 
propagation distance.  Colors show logarithmic scale emittance growth 
and 1\% and 10\% threshold contours (dashed) are labeled 
separately.  The extent of the envelope instability band 
and Tiefenback's stability threshold are indicated.  
Results for initial 
waterbag and thermal distributions are similar, but the transition to 
instability has more structure for the waterbag distribution.  
Strong growth regions in all cases 
are qualitatively consistent with Tiefenback's threshold.   
Emittance growth cannot be attributed to KV-like modes internal to 
the beam\cite{Hofmann-1983,Lund-1998}.  Much of
the emittance growth is associated with particles that evolve 
significantly outside the 
core (see Fig.~\ref{Fig:sim-emit-evolve}b) rendering any linear internal 
mode interpretation questionable.  Also, 
many KV modes are strongly unstable 
(instabilities exist for $\sigma/\sigma_0 < 0.3985$ even in the 
continuous focusing limit) where no rms emittance 
growth is observed.  KV modes generally predict wrong parametric 
variations of instability (thresholds bend the wrong way).  
Large internal modes also possess little free energy to drive 
statistical emittance growth\cite{Reiser-1994} and therefore may 
not be dangerous if they saturate at small amplitudes.  

\section{Core-Particle model} 

Consider an unbunched beam of ions of charge $q$ and mass $m$
propagating with axial velocity $\beta_b c$ ($c$ is the speed of
light {\em in vacuuo}) and relativistic factor $\gamma_b = 1/\sqrt{1
-\beta_b^2}$.   A linear applied focusing lattice is assumed,
self-field interactions are electrostatic.  Then the transverse
orbit $x(s)$ of a beam particle evolves according to the paraxial 
equations of motion\cite{Lund-2004,Reiser-1994}
\begin{equation}
x'' + \kappa_x x  = - \frac{q}{m\gamma_b^3\beta_b^2 c^2}
                     \frac{\partial\phi}{\partial x} .
\label{Eq:part-eqns}
\end{equation}
Here, $s$ is the axial coordinate of a beam slice, primes denote
derivatives with respect to $s$, and $\kappa_x(s)$
is the linear applied focusing function of the lattice (specific
forms can be found in Ref.~\cite{Lund-2004}),
and the electrostatic potential $\phi$ is related to the number
density of beam particles $n$ by the Poisson equation
$\nabla_\perp^2 \phi = -q n/\epsilon_0$ in free-space.
$\epsilon_0$ is the permittivity
of free space.

The core of the beam is centered on-axis ($x= 0 = y$),
and is uniform density within an elliptical cross-section with edge
radii $r_j$ (henceforth, $j$ ranges over both $x$ and $y$)
that obey the KV envelope equations
\begin{equation}
r_j'' + \kappa_j r_j - \frac{2Q}{r_x + r_y} - \frac{\emit_j^2}{r_j^3} = 0 .
\label{Eq:KV-env-eqns}
\end{equation}
Here, $Q = q\lambda/(2\pi\epsilon_0 m\gamma_b^3 \beta_b^2 c^2) = \const$
is the dimensionless perveance,
($\lambda = q n(x=0,y=0) r_x r_y = \const$ is the beam
line-charge density), and $\emit_j$ is the
rms edge emittance along the $j$-plane.  We
take $\emit_j \equiv \emit  = \const$.   For a periodic
focusing channel with lattice period $L_p$,
$\kappa_j(s + L_p) = \kappa_j(s)$, the envelope is called matched
when it has the periodicity of the lattice, i.e., 
$r_j(s+L_p) = r_j(s)$.  Undepressed particle phase advances are
used to set the lattice focusing functions $\kappa_j$ using
$\cos\sigma_0 = (1/2){\rm Tr}\;{\bf M}$ where ${\bf M}$ is the $x$ or
$y$ plane transfer map of a single particle ($Q = 0$) through
one lattice period.  We take the $\kappa_j$ to be piecewise constant 
with occupancy $\eta \in (0,1]$.  The matched beam envelope flutter 
varies only weakly with $\eta$ but increases strongly with increasing 
$\sigma_0$.  The depressed particle phase advance is calculated
as $\sigma = \emit\int_0^{L_p}\! ds/r_j^2$.


It can be shown that the flutter
of the matched beam envelope for periodic FODO
quadrupole focusing systems with piecewise
constant $\kappa_j(s)$ is 
given approximately (for $\sigma/\sigma_0 \ll 1$) by\cite{Lee-2002}
\begin{equation}
\frac{ r_x|_{\rm max}}{\bar{r_x}} -1 \simeq
  (1-\cos\sigma_0)^{1/2}\frac{(1-\eta/2)}{2^{3/2}(1-2\eta/3)^{1/2}}
\label{Eq:flutter}
\end{equation}
Here, $\eta \in (0,1]$ is the occupancy of the quadrupoles in the
lattice and $\bar{r_x} = (1/L_p)\int_0^{L_p}\! ds \; r_x$.
Equation~(\ref{Eq:flutter}) shows that
envelope flutter in a quadrupole channel depends strongly on
$\sigma_0$ and weakly on $\eta$
(the variation in $r_x|_{\rm max}/\bar{r_x}$ in $\eta$ is $\sim 13\%$).

For a particle evolving both inside and outside the
elliptical beam envelope, Eq.~(\ref{Eq:part-eqns}) can be expressed as
\begin{equation}
x'' + \kappa_x x  = \frac{2Q F_x}{(r_x + r_y)r_x}x ,
\label{Eq:core-part-eqns}
\end{equation}
with an analogous equation for the $y$-plane.
Here, $F_j$ are form factors satisfying $F_j = 1$ inside the beam
($x^2/r_x^2 + y^2/r_y^2 \leq 1$) and
$F_x  =  (r_x + r_y)\frac{r_x}{x}{\rm Re}[ \underline{S} ]$ and
$F_y  = -(r_x + r_y)\frac{r_y}{y}{\rm Im}[ \underline{S} ]$
outside the beam ($x^2/r_x^2 + y^2/r_y^2 > 1$).
$\underline{S}$ is a complex variable defined as
$\underline{S}  \equiv \frac{\underline{z}}{(r_x^2 - r_y^2)}[
1 - \sqrt{ 1 - \frac{r_x^2 - r_y^2}{\underline{z}^2} } ]$,  where
$\underline{z} = x + i y$ and $i = \sqrt{-1}$.

The particle equations of motion~(\ref{Eq:core-part-eqns})
are integrated numerically from initial
conditions.
We typically launch particles with initial
$x$ and $y$ coordinates outside the beam edge
(i.e., $x^2/r_x^2 + y^2/r_y^2 > 1$) and with initial angles
$x'$ and $y'$ consistent with coherent flutter motion of core particles
extrapolated to the location of the particle, i.e., with
$x' = r_x' x/r_x$ and $y' = r_y' y/r_y$.
Diagnostics include particle trajectories, single particle
emittances defined by
$\epsilon_x = \sqrt{ (x/r_x )^2 + ( x r_x' - x' r_x)^2/\emit_x^2 }$
($\epsilon_x = 1$ at the core distribution edge), stroboscopic
Poincare phase space plots, and particle oscillation
wavelengths calculated from Fourier transforms of orbits.
Particle trajectories and phase-spaces
analyzed in scaled units (e.g., with $x$--$x'$ projections
scaled as $x/r_x$--$(x' r_x - r_x' x)/\emit_x$) to better illustrate
oscillation extents relative to the matched beam core.

\section{Core-Particle simulations }

\begin{figure}[htb]
\centering
\includegraphics*[width=75mm]{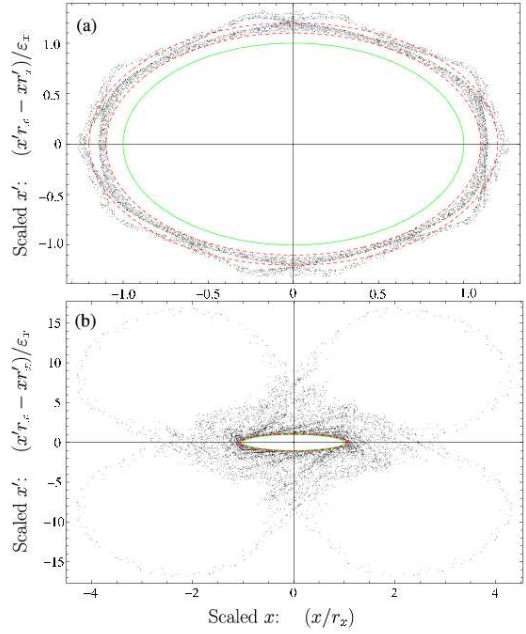}
\caption{
(color online) Core-particle Poincare phase-spaces for
$\sigma_0 = 100^\circ$, $\sigma/\sigma_0 = 0.67$ (a), and
$\sigma/\sigma_0 = 0.2$ (b).  ($L_p = 0.5$ m, $\eta = 0.5$,
$\emit = 50$ mm-mrad). 
}
\label{Fig:core-part-sim}
\end{figure}

To  illustrate the halo structure, we launch particles along the
$x$-axis of the elliptical beam in specified regions
outside the beam edge (e.g., $x \in [1.1, 1.2] r_x$) with zero
incoherent angle spreads (e.g., $x' = r_x' x/r_x$).
Fig.~\ref{Fig:core-part-sim} illustrates $x$--$x'$ Poincare
phase-spaces for particles launched 
with $ x \in [1.1,1.2] r_x$ for fixed
$\sigma_0$ and two values of $\sigma/\sigma_0$: (a) a high value
(weak space-charge) well within the stable region of
Fig.~\ref{Fig:transport-params}, and (b) a low value (strong space
charge) in the unstable region.  The Poincare strobe is one lattice
period.  Scaled coordinates $x/r_x$ and $(x' r_x - x r_x')/\emit_x$
are plotted to remove envelope flutter.   The extent of the core is
plotted in red.  Extrapolations of the range of initial launch conditions
are indicated in red based on the annular elliptical region formed if the
initial particle conditions evolved with constant single-particle
emittance $\epsilon_x$.  Note the large change in scale between the 
stable and unstable plots. For the stable case, 
particles diving in and out of the
matched envelope remain close to the initial launch range and
indicate a weak, high-order resonance.  For the unstable case,
numerous resonances near the core become stronger and overlap
causing the region immediately 
outside the core to break up into a stochastic
sea that closely approaches the core.   A large, $4$-lobe bounding 
resonance (KAM surface) persists that
ultimately limits the achievable particle oscillation amplitude.   
The phase advance of particles moving outside the envelope is strongly 
amplitude dependent ranging from $\sigma$ for amplitudes at the core boundary 
to $\sigma_0$ for very large amplitudes.  Strong space charge 
($\sigma/\sigma_0  \ll 1$) and large matched envelope 
oscillations (large $\sigma_0$) provide a strong pump at the lattice 
frequency.  Numerous harmonics of  
particle orbits near the core resonate with the lattice 
resulting in overlapping resonances that produce a strongly chaotic 
region that approaches the core.  This chaotic sea allows 
particles near the core to rapidly evolve to large amplitudes.  


%

A new stability criterion is adopted to estimate where chaotic halo orbits 
near the beam core can degrade transport.   When 
varying $\sigma_0$ and $\sigma/\sigma_0$, we define the stability 
boundary to be the first point when approached from stable regions (low 
$\sigma_0$) where particle groups launched near the core (e.g., 
$x \in [1.05,1.10]r_x$) experience large increases in 
oscillation amplitude (e.g., Max$[x/r_x]$ increased to $1.5$).    
Boundary points obtained when particles launched with 
$x/r_x \in [1.05,1.10]$ increase in amplitude by factors of 
$1.5$ (triangles) and $1.4$ 
(squares) are plotted in Fig.~\ref{Fig:transport-params}.  The boundary 
roughly tracks the region of strong emittance growth observed in 
experiment and simulations until the envelope instability band is approached.  
Results are relatively insensitive to the choice in initial group radius 
and amplitude increase factor.  Earlier work by Langiel\cite{Lagniel-1994} 
employed a core-particle model to analyze transport limits but implied 
overly pessimistic stability criteria ($\sigma_0 < 60^\circ$ and 
$\sigma/\sigma_0 > 0.4$) seemingly based on rough 
resonance overlap estimates.   

%
\begin{figure}[htb]
\centering
\includegraphics*[width=80mm]{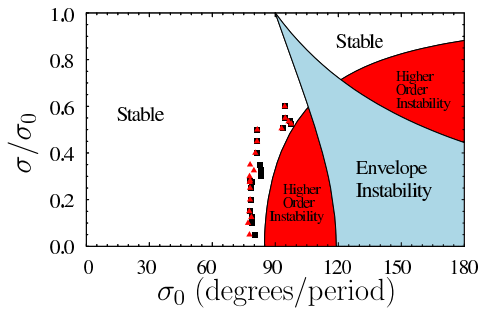}
\caption{
(color online) Beam stability boundary calculated from a core-particle 
model for a FODO quadrupole channel.  ($L_p = 0.5$ m, $\eta = 0.5$,
$\emit = 50$ mm-mrad). 
}
\label{Fig:core-part-stab-bound}
\end{figure}

Halo properties analyzed persist when particles have 
finite angular momentum (not launched on-axis).  
Particles that leave the core in self-consistent 
PIC simulations generate similar Poincare plots to the 
core-particle model for a variety of initial distributions.   
Single particle emittance
growths of $\sim 50$ are possible for particles near the beam edge 
that enter the halo in unstable regions.  If a significant number 
of near-edge particles enter the halo, this can result in strong 
increases in rms beam emittance and distortions in the beam 
phase-space (both total and core).  Particles 
leaving the core in unstable regions 
rapidly grow in amplitude over a relatively small
number of lattice periods -- consistent
with PIC simulations.  Moreover, as observed in simulations and 
experiment, this halo induced mechanism for transport
degradation is consistent with increasing sensitivity to the beam
distribution and edge perturbations as the threshold region
is approached.   
The core-particle model assumption of a uniform density 
elliptical beam core is reasonable for strong space-charge due to
Debye screening and phase-mixing of initial perturbations.  
No periodic, nonuniform density equilibria
are known and core perturbations are observed in PIC simulations to
collectively evolve and disperse leaving smaller residual fluctuations and
a rounded beam edge.  Hence the uniform core model can provide 
a good approximation to the average impulse a halo particle experiences
while traveling through the oscillating core.
If the edge of the beam distribution is not sharp, as is expected
for finite $\sigma/\sigma_0$, a significant population  of
edge particles can enter the halo and be elevated to large amplitudes in 
unstable regions.  Due to envelope flutter,
the spatial average temperature $\bar{T_x}$ of a beam with constant
emittance $\emit_x$ will vary as $\bar{T_x} \propto \emit_x^2/r_x^2$.
Thus, in a matched beam envelope the temperature will
oscillate with the period of the lattice
($360^\circ$ phase advance), decreasing where the envelope contracts and
conversely decreasing where the envelope expands.  For high $\sigma_0$ these
fluctuations will tend to increase leaving the edge out of force-balance.
On the other hand,  the plasma response of the beam will have
characteristic collective phase advance
$\sigma_p = \frac{180^\circ}{\pi}\frac{L_p}{r_x}\sqrt{2Q}$
This frequency will generally be incommensurate with and much slower than the
temperature oscillations showing showing that the edge of the beam will
have a more difficult time readapting to the focusing kicks as
$\sigma_0$ increase and envelope flutter becomes larger.

\section{Conclusions}

A core-particle model was used to analyze the previously unexplained 
origin of space-charge related transport limits of beams propagating in 
periodic quadrupole focusing channels.  
It was shown that when matched beam envelope oscillations 
and space-charge strength are both sufficiently large, 
near-edge particles oscillating both inside and outside the matched
beam envelope become chaotic and can experience large increases in 
oscillation amplitude.  This resonance halo need not be tenuous and 
is distinct from envelope mismatch driven halo\cite{Gluckstern-1994}
because the driving oscillation
is the fast flutter of the matched beam envelope rather than envelope 
mismatch modes.  The matched
envelope flutter becomes larger with increasing $\sigma_0$, providing a
strong pump that further increases as beam space-charge forces 
become larger.  Envelope oscillations also drive large temperature
oscillations in the core of the matched envelope.  Because the collective
response of the beam to local force imbalances
scales with the plasma frequency
which is low relative to the lattice frequency, it is unlikely that the beam
edge can consistently adapt.  Lack of edge 
self-consistency in periodically focused beam distributions 
makes it plausible that many near-edge
particles can move sufficiently outside the beam core to partake in the
resonance.   Consequently, large distortions in the beam phase-space and
large rms emittance growth can result.  
Stability thresholds based on this resonance picture 
are in rough agreement with experimental measurements and simulations.
Analogous transport limits to the ones studied here will occur in other 
periodic focusing channels. Generally, unstable parameters will differ 
due to different scaling 
of matched beam envelope flutter.  Envelope mismatch also increases 
driving envelope excursions and introduces additional frequencies -- likely 
reducing the region of stable transport.  
Work is ongoing to further clarify the processes described.  Further details 
of this work can be obtained on the arXiv e-print 
server\cite{Lund-2006} and in future publications.   


\section*{Acknowledgments}

B. Bukh and J. Barnard helped develop parts of the core-particle 
model.  D. Grote aided the WARP simulations.  J. Barnard,
I. Haber, E. Lee, and P. Seidl provided useful discussions.  This
research was performed at LLNL and LBNL under US DOE contact
Nos.~W-7405-Eng-48 and DE-AC03-76SF0098.


\begin{thebibliography}{9}
\bibitem{Tiefenback-1985} M.G. Tiefenback and D. Keffe, IEEE 
  Trans.\ Nuc.\ Sci., NS-32, 2483 (1985).    
\bibitem{Tiefenback-1986} M.G. Tiefenback, 
  {\em Space-Charge Limits on the Transport of Ion Beams}, 
  U.C. Berkeley Ph.\ D thesis and Lawrence Berkeley Lab report 
  LBL-22465 (1986).
\bibitem{Haber-misc} Simulations were carried out by 
  Haber, Laslett, and colleagues that first suggested the limit.  Partial  
  results were reported in: I. Haber, IEEE Trans.\ Nucl.\ Sci.\ {\bf NS-26}
  3090 (1979);  I. Haber and A.W. Maschke, Phys.\ Rev.\ Lett.\ {\bf 42}, 
  1479 (1979).  
\bibitem{Struckmeier-1984} J. Struckmeier, J. Klabunde, and M. Reiser, 
  Particle Accel.\ {\bf 15}, 47 (1984).   
\bibitem{Maschke-1972} A.W. Maschke, {\em Heavy Ion Space Charge Limits,} 
  Technical Report BNL 20297, Brookhaven National Laboratory, July 1975.  
\bibitem{Lund-2005} S.M. Lund and S.R. Chawla, {\em Space-Charge Transport 
  Limits in Periodic Channels,} Proceedings of the 2005 
  Particle Accelerator Conference, Knoxville TN, paper FPAP034.  
\bibitem{Lund-2004} S.M. Lund and B. Bukh, PRSTAB {\bf 7} 024801 (2004).
\bibitem{Reiser-1994} M. Reiser, {\em Theory and Design of
  Charged Particle Beams,} (Wiley, 1994), and references therein.
\bibitem{Grote-1998} D.P. Grote, A. Friedman, I. Haber, W. Fawley, and 
  J.-L. Vay, Nuc.\ Instr.\ Meth.\ A {\bf 415} 428 (1998).
\bibitem{Hofmann-1983} I. Hofmann, L.J. Laslett, L. Smith, and I. Haber, 
  Particle Accel.\ {\bf 13}, 145 (1983).      
\bibitem{Lund-1998} S.M. Lund and R.C. Davidson, Phys.\ Plasmas {\bf 5}, 
  3028 (1998).     
\bibitem{Lee-2002} E. P. Lee, Phys.\ Plasmas {\bf 9}, 4301 (2002). 
\bibitem{Lagniel-1994} J.-M.\ Lagniel, Nuc.\ Instr.\ Meth. A, {\bf 345} 
  405 (1994).  
\bibitem{Gluckstern-1994} R.L. Gluckstern, 
  Phys.\ Rev.\ Lett. {\bf 73}, 1247 (1994).
\bibitem{Lund-2006} S.M. Lund and S.R. Chawla, to be posted on 
  http://www.arxiv.org.  

\end{thebibliography}
\end{document}